\documentclass[prl,aps,
nofootinbib,  
superscriptaddress,showpacs,twocolumn,longbibliography]{revtex4-1}

\usepackage{graphicx} 
\usepackage[colorlinks]{hyperref}
\usepackage{amsmath,amssymb, physics} 
\usepackage[dvipsnames]{xcolor}

\usepackage{bm} 

\usepackage{bbold}  

\newcommand{\beq}{\begin{equation}}
\newcommand{\eeq}{\end{equation}}

\usepackage{mathtools}
\usepackage{comment}
\usepackage{amsmath} 

\hypersetup{linkcolor=magenta,
    citecolor=blue}

\begin{document}

\title{When does nonreciprocity matter? Scale-dependence and nonequilibrium signatures}

\author{Sarah A.M. Loos}
\affiliation{Max Planck Institute for Dynamics and Self-Organization, G\"ottingen, Germany}
\author{Robert L. Jack}
\affiliation{Department of Applied Mathematics and Theoretical Physics, University of Cambridge, Wilberforce Road, Cambridge CB3 0WA, United Kingdom}
\affiliation{Yusuf Hamied Department of Chemistry, University of Cambridge,
Lensfield Road, Cambridge CB2 1EW, United Kingdom}

\begin{abstract}
Nonreciprocity is increasingly recognized as a unifying concept linking diverse nonequilibrium phenomena found across physics, chemistry, and biology.
It gives rise to distinctive behavior including run-and-chase dynamics and spatio-temporal patterns, often associated with a breaking of time-reversal symmetry. However, nonreciprocity and its nonequilibrium signatures are fundamentally scale-dependent, and may emerge or disappear under coarse-graining. A central challenge is therefore to understand when and how nonreciprocity manifests itself on different scales, for example via irreversible fluctuations or macroscopic currents.  In this Perspective, we discuss the physical origin and fate of effective nonreciprocal interactions and the characteristic irreversible dynamics they give rise to across scales.
\end{abstract}

\maketitle

Classical phases of matter are shaped by the interactions between their constituent atoms and molecules. These interactions are \emph{reciprocal}, meaning that they obey the action-reaction principle: the force on particle $i$ from particle $j$ is equal and opposite to the force on $j$ from $i$. This fundamental principle derives from conservation of momentum, which is deeply embedded in physical theories.  

Effective \emph{nonreciprocal} (NR)
interactions~\cite{fruchart2026nonreciprocal}, which violate the action-reaction principle, 
nevertheless arise naturally 
in nonequilibrium systems, where the particles' dissipative environment can absorb or \textit{supply} momentum. Well-studied examples include 
complex dusty plasmas~\cite{ivlev2015statistical}, 
colloidal particles whose interactions are mediated by chemical fields~\cite{soto2014self,dinelli2023non,agudo2019active}, and mixtures of active and passive particles~\cite{stenhammar2015activity,you2020nonreciprocity,mason2025dynamical}. 
Such interactions generate qualitatively new phenomena, including run-and-chase dynamics, which can set a two-body system into spontaneous motion, macroscale spatiotemporal patterns not seen in equilibrium, like traveling density waves, oscillatory phases~\cite{fruchart2021non,saha2020scalar,you2020nonreciprocity,suchanek2023irreversible,suchanek2023entropy,alston2023irreversibility}, as well as new classes of (universal) critical behaviour~\cite{fruchart2026nonreciprocal,zelle2024universal,johnsrud2026scaling}.

\begin{figure*}
\centering
\includegraphics[width=17cm]{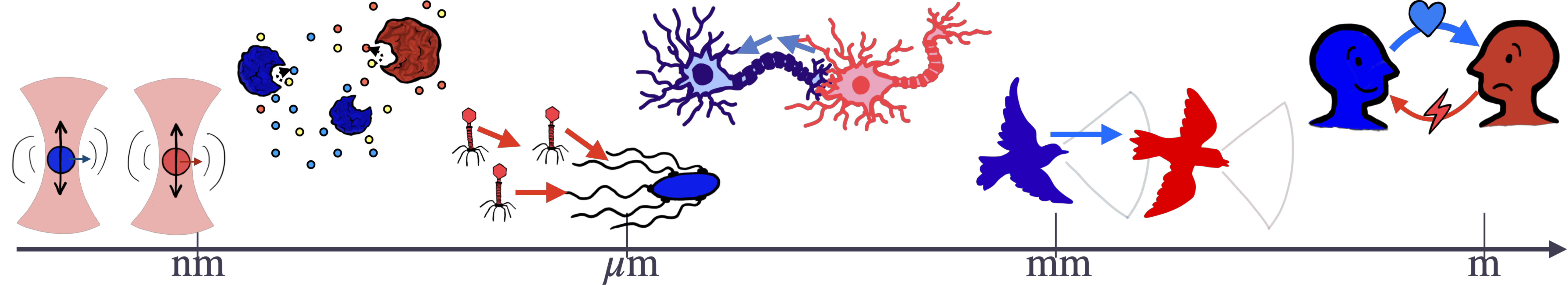}
\caption{Effective nonreciprocal interactions arise in a wide variety of systems, spanning many length scales. Examples include optomechanical levitated nanoparticles, where interference can be used to engineer tunable nonreciprocal couplings~\cite{rieser2022tunable}; mixtures of catalytically active and chemophoretic enzymes~\cite{jee2018catalytic,soto2014self,dinelli2023non,agudo2019active};
phage--bacteria systems exhibiting run-and-chase dynamics; neuronal networks with directed information flow; animal and robotic systems interacting through directed perception~\cite{dadhichi2020nonmutual,loos2023long}; and social systems with asymmetric preference or influence~\cite{strogatz2024nonlinear}. 
Despite their diverse characteristics, these systems share interaction asymmetry, motivating the search for universal principles governing NR matter. 
} \label{Fig1}
\end{figure*}

Interaction asymmetry is also widespread beyond physical systems.  It naturally appears in ecology 
as predator--prey and host--parasite interactions;  in
neural circuits as asymmetric synaptic couplings; 
in social systems through asymmetric influence; and in economic or game-theoretic multi-agent settings through conflicting individual objectives~\cite{strogatz2024nonlinear}. Some examples are given in Fig.~\ref{Fig1}. Nonreciprocity is also increasingly exploited as a design principle for active metamaterials and robotic systems with programmable functionalities~\cite{brandenbourger2019non}.

Despite their disparate microscopic origins, NR interactions lead to remarkably similar collective dynamics across these systems, including oscillatory and travelling macroscopic states.  
At continuum level, such behavior is frequently captured by universal, deterministic models: the complex Ginzburg-Landau equation arises across physics, chemistry, biology, and ecology~\cite{aranson2002world}; the NR Cahn-Hilliard (NRCH) equation has recently been established as a prototypical theory for NR systems with particle number conservation~\cite{you2020nonreciprocity,saha2020scalar,frohoff2023nonreciprocal}. At the microscopic level, nonreciprocal spin and lattice models provide minimal statistical-mechanical descriptions that incorporate fluctuations and stochasticity, ranging from finite-dimensional models~\cite{avni2025nonreciprocal,loos2023long,dadhichi2020nonmutual} to mean-field spin-glasses with asymmetric couplings~\cite{sompolinsky1988chaos,garcia2025nonreciprocal}. 

Since the notion of nonreciprocity extends far beyond particle mechanics, it cannot always be tied back to effective breaking of momentum conservation. However, reciprocity also has another aspect: 
it is associated with the existence of a \textit{common} interaction potential.
In contrast, NR  dynamics instead follows gradients of individual objective functions, which have been called ``selfish energies''~\cite{avni2025nonreciprocal}, reflecting different (possibly conflicting) desires among the constituents.
This viewpoint naturally encompasses reaction--diffusion systems, ecological and evolutionary models, neural networks, and socioeconomic systems, whose dynamics is often governed by optimization principles; analogous to the minimization of a thermodynamic potential in physics.
Similarly, in macroscopic descriptions, where the relevant degrees of freedom are order parameters $\phi_1,\phi_2,\dots$ (like chemical concentrations, population densities, magnetizations), the dynamics may follow the gradient of a free energy $F$, and 
reciprocity of interactions derives from
 $\frac{\partial^2F}{\partial \phi_i\partial \phi_j} = \frac{\partial^2F}{\partial \phi_j\partial \phi_i}$. NR dynamics can then be understood as following gradients of selfish free energies, which are different for each order parameter.

The notion of selfish (free) energy also highlights the intimate connection of nonreciprocity to nonequilibrium statistical mechanics.
Selfish energies encode competition for different objectives that can sustain persistent (mass) currents; rather than relaxing towards a common state of satisfaction (energetic minimum). Consequentially, many characteristic signatures of NR systems, like sustained oscillations, travelling waves, and limit cycles despite damping, 
are associated with breaking of \textit{time-reversal symmetry} (TRS), even in their steady states, which reflects ongoing underlying dissipation~\cite{seifert2012stochastic}. 

In stochastic models, the degree of TRS breaking can be quantified by irreversibility measures such as the entropy production rate, that provides a natural framework to formally address scale-dependent irreversible phenomena associated with nonreciprocity \cite{
zhang2023entropy,loos2020irreversibility,loos2023nonreciprocal,suchanek2023irreversible,suchanek2023entropy,alston2023irreversibility,dopierala2026odd}.
Interestingly, such measures tend to be reduced under coarse-graining, as parts of the dissipative probability currents can get hidden at larger scales~\cite{esposito2012stochastic,kawaguchi2013fluctuation}. 
This observation motivates one of the central themes of this Perspective: while nonreciprocity may emerge, disappear, or become hidden across levels of description, its nonequilibrium signatures evolve just as nontrivially. Understanding the relationship between nonreciprocity and TRS is therefore essential for connecting interaction asymmetries to associated irreversible phenomena across scales.

\textit{Scope.} This Perspective considers nonreciprocity on two complementary levels of description. The first comprises mesoscale models, in which particles (or spins) interact through effective forces that violate action-reaction symmetry. The second comprises macroscopic continuum field theories, in which different fields evolve according to gradients of distinct selfish free energies rather than a common thermodynamic potential.
 We emphasize that the relation between these descriptions is highly nontrivial: Reciprocal particle interactions can generate effective nonreciprocal continuum theories, as found in active--passive mixtures~\cite{stenhammar2015activity,you2020nonreciprocity}, but mesoscale nonreciprocity may also become hidden or disappear under coarse-graining. 
This raises two fundamental questions of ongoing interest: When do NR interactions survive coarse-graining and lead to macroscopic consequences? And what can be concluded from macroscopic NR signatures about the underlying interactions and dissipation?

Beyond the classes of systems considered here, numerous mathematical analogies can be drawn between interaction nonreciprocity and other types of asymmetric or non-Hermitian dynamics, including odd elasticity~\cite{scheibner2020odd}, nonreciprocal optics, and gyroscopic systems~\cite{fruchart2026nonreciprocal}. However, we  restrict our discussion primarily to interaction asymmetry between distinct physical entities, rather than couplings between different degrees of freedom of the same object or nonreciprocal wave propagation.  These may result in similar dynamical features, but they have a fundamentally different physical interpretation.
(For example, Hamiltonian dynamics involves asymmetric coupling between momentum $p$ and position $x$, yet we do not regard this as nonreciprocity.)

\textit{TRS and Entropy Production.}
Since TRS breaking plays a central role throughout this Perspective, we briefly recall its definition.
Consider the dynamics
$
\dot{\mathbf z}
=
\mathbf G(\mathbf z)
+\boldsymbol{\eta},
$ with deterministic drift $\mathbf G$ and stochastic forcing $\boldsymbol{\eta}$,
where the components of $\mathbf z=(z_1,\dots,z_n)$ may be scalars, $z_i\in \mathbb{R}$, or (order parameter) fields, $z_i(x,t)\in \mathbb{R}$; and have either even
($\epsilon_i=1$) or odd ($\epsilon_i=-1$) parity under time reversal operation $\Theta\mathbf z=(\epsilon_1z_1,\ldots,\epsilon_nz_n)$.

The dynamics obeys TRS if every trajectory
$\Gamma=\{\mathbf z(t)\}_{0\le t\le\tau}$
is has the same path probability as its time-reversed counterpart
$\Theta\Gamma=\{\Theta\mathbf z(\tau-t)\}$,
\begin{equation}
\mathbb P[\Gamma]
=
\mathbb P[\Theta\Gamma].
\label{eq:TRS}
\end{equation}
A quantitative measure of TRS breaking is the stochastic (trajectory-wise) entropy production rate~\cite{seifert2012stochastic,esposito2012stochastic,kawaguchi2013fluctuation},
\begin{equation}
\Sigma[\Gamma]
=
\lim_{\tau\to0}
\frac{k_{\rm B}}{\tau}
\ln
\frac{\mathbb P[\Gamma]}
{\mathbb P[\Theta\Gamma]},
\label{eq:epr_general}
\end{equation}
whose steady-state average is non-negative and vanishes if and only if detailed balance holds.

\section{Nonreciprocal particle dynamics: from micro- to meso-scale}
Physical systems are naturally described at different levels of resolution. For microscopic descriptions that include all relevant degrees of freedom, the complete  dynamics conserves total momentum, implying that interaction forces between any microscopic constituents $(i,j)$ satisfy action-reaction symmetry $F_{ij}=-F_{ij}$, even out of equilibrium. 
 
In contrast, in mesoscale descriptions some fast or inaccessible degrees of freedom have been eliminated, for example by coarse-graining. Then
 the resulting pairwise coupling forces may violate action--reaction symmetry,
$F_{ij} \neq-F_{ji}$~\cite{ivlev2015statistical}. In particle systems, 
this can happen, e.g., through elimination of a nonequilibrium environment that mediates the interactions \cite{dinelli2023non,soto2014self,agudo2019active}, such as chemical, hydrodynamic, optical or plasma fields. While the original system ``particles plus environment'' still obeys microscopic momentum
conservation, the reduced particle dynamics generally does not, because momentum is continuously exchanged with the eliminated environment.
Nonreciprocity in physical systems should therefore be regarded
as an emergent property of coarse-grained, reduced, or effective dynamics; rather than a
fundamental violation of microscopic mechanics.

A prime example appears in mixtures of particles that are both catalytically active and chemophoretic (or chemotactic); properties found, e.g., in enzymes~\cite{jee2018catalytic}, catalytic colloids, chemotactic cells~\cite{van2004chemotaxis}, and bacteria~\cite{wadhams2004making} (Fig.~\ref{Fig2}). Catalytic activity generates chemical gradients, while chemophoresis biases motion along them; naturally generating NR interactions. If species A is attracted to the chemical trails of species B while species B is repelled by those of A, this may even produce \textit{antagonistic} coupling ($F_{ij}F_{ji}>0$) leading to run-and-chase dynamics. Formally, NR interactions appear in the mesoscale description as traces of the nonequilibrium character of the eliminated chemical field.

In thermal equilibrium, eliminating degrees of freedom preserves the existence of a common thermodynamic potential (effective free energy, potential of mean force) governing the reduced dynamics. 
Although friction, noise, and memory may emerge, the reduced dynamics continues to satisfy detailed balance, and relaxes towards the corresponding (Gibbs) equilibrium distribution. Since all effective interaction forces derive from the common potential, they remain reciprocal. Nonreciprocity is therefore generically a nonequilibrium phenomenon. 
However, it alone does not determine whether the dynamics is (ir)reversible, which depends on both the deterministic drift and the stochastic forcing.

\textit{Example of NR and TRS.}
To discuss the connections between NR, TRS and broken momentum conservation, it is instructive to consider a paradigmatic mesoscale example, for which we use the Markovian Langevin equation,
\begin{equation}
m_i\ddot X_i
=
\sum_j F_{ij}
-\gamma_i \dot X_i+
\sqrt{2\gamma_i k_\mathrm{B} T_i} \,
\nu_i ,
\label{eq:underdamped}
\end{equation}
where 
$\nu_i$ is a zero-mean Gaussian white noise with 
$
\left\langle \nu_i(t)\nu_j(t')\right\rangle
=
\delta_{ij}\delta(t-t')$.
The coordinate $X_i$ represents the position of a particle $i$ with mass $m_i$, and is even under time reversal, $(\Theta X_i, \Theta \dot X_i)=(X_i,-\dot X_i)$. The friction term with coefficient $\gamma_i$ and the noise both stem from coupling to a bath, i.e., a collection of faster degrees of freedom that have been eliminated from the description. 
In this physical picture, $m_i \dot X_i$ is the momentum of particle $i$ and $F_{ij}$ act as physical interaction forces. 
We write $M=\sum_i m_i$ for the total mass.

For the reciprocal case ($F_{ij}=F_{ij}$), the center of mass $\sum_i m_i X_i/M$ performs an Ornstein-Uhlenbeck process, where frictional damping $\sum_i \gamma_i \dot X_i$ acts as a momentum sink (flux of momentum to the bath), and the noise  describes stochastic 
exchange of momentum and energy between bath and system, that is zero on average. Thus, while the coupling to the bath breaks momentum conservation, there is no net momentum injection. Nevertheless, systems with $T_i\not = T_j$ are out of equilibrium, and TRS is generically broken.

In contrast, the NR case means that the center of mass is driven by $\sum_{ij} F_{ij}$, and there exist configurations for which the particle subsystem experiences systematic \textit{injection of linear momentum} from its surrounding. This becomes particularly obvious in systems with run-and-chase dynamics. 

But, does nonreciprocity \textit{necessarily} lead to a breaking of TRS at the mesoscale?
To discuss this, we restrict for simplicity to linear forces, $F_{ij}=A_{ij}(X_j-X_i)$,
where nonreciprocity simply means $A_{ij}\neq A_{ji}$.
The corresponding particle-wise selfish energy  $
    E_i = \frac{1}{2} \sum_j A_{ij}(X_i-X_j)^2$, reduces in the reciprocal limit to the global 
 interaction potential $V = \frac{1}{2} \sum_iE_i$, and Eq.~\eqref{eq:underdamped}
describes a thermal diffusion in a parabolic potential.

In the NR case, the stationary dynamics 
features non-vanishing probability currents, and the entropy production rate \eqref{eq:epr_general} scales as~\cite{loos2020irreversibility,loos2023nonreciprocal}
\begin{equation}
\Sigma
\propto \left(A_{ji}T_i-A_{ij}T_j\right)^2 \geq 0\,,
\label{eq:epr}
\end{equation}
which vanishes, if for all pairs $(ij)$
\begin{equation}
A_{ij}T_j=A_{ji}T_i.
\label{eq:DB}
\end{equation}
For this class of linear Langevin systems, the remarkably simple condition \eqref{eq:DB} is necessary and sufficient for TRS and detailed
balance~\cite{loos2020irreversibility,loos2023nonreciprocal}. 

The condition (\ref{eq:DB}) provides several insights into the
relation between nonreciprocity and TRS at linear order. 
First, unless very carefully tuned, NR generically breaks TRS.
Second, Eq.~(\ref{eq:DB}) is independent of the masses, friction,
and the diagonal components of the drift: these quantities affect relaxation speeds and the magnitude of irreversible currents, but not whether the stationary
dynamics is irreversible or whether it is nonreciprocal.
Third, Eq.~\eqref{eq:DB} shows that asymmetric couplings can, under certain conditions, be exactly compensated
by unequal bath temperatures. We note that this is also possible for nonlinear interactions that share the same spatial dependencies~\cite{zhang2023entropy}. 
{Furthermore, for certain types of distance-independent NR interactions, similar ``compensations of nonreciprocity'' by thermodynamic drivings have also been formalized in~\cite{ivlev2015statistical} where appropriately renormalized masses and interaction potentials yield a pseudo-Hamiltonian description; also see~\cite{shi2026hamiltonian}.

\textit{Can nonreciprocity be deduced from particle trajectories?}
We have introduced nonreciprocity as a property of interactions and emphasized its implications for the resulting dynamics.  
However, given some experimental data, inferring the underlying interaction forces is generally a nontrivial problem, and often hinges on prior physical knowledge.
There  are even cases where reciprocal and NR systems exhibit identical dynamics (an example is given below).
While this might suggest that nonreciprocity is ambiguous, our viewpoint is instead that it is a well-defined \textit{physical property} of the underlying system, even if it cannot always be inferred uniquely from the observed dynamics or equations of motion, alone.

For a specific example, consider Eq.~\eqref{eq:underdamped} and define
 $\tilde{m_i}=m_i \mu_i$, 
$\tilde A_{ij}=A_{ij} \mu_i$,
$\tilde\gamma_i= \gamma_i \mu_i$, $\tilde T_i=T_i\mu_i$ with some $\mu_i>0$.
Multiplying through by $\mu_i$, this Eq.~\eqref{eq:underdamped} becomes
\begin{align}\label{eq:mapped-dynamics}
\tilde{m}_i \ddot X_i = 
\sum_j \tilde{A}_{ij} (X_j-X_i)-\tilde{\gamma}_i \dot X_i
+ \sqrt{2 \tilde{\gamma_i}k_\mathrm{B}\tilde{T}_i } \,\eta_i .
\end{align}
If we now interpret $\{\tilde m_i, \tilde T_i,\tilde \gamma_i,\tilde A_{ij}\}$ as \textit{physical} masses, temperatures, friction and interaction coefficients,
Eq.~\eqref{eq:mapped-dynamics} describes a different physical system from the original one: the two systems have identical dynamics, but different masses, temperatures, and forces.
Moreover, if 
\begin{align}\label{condition2}
   \mu_i / \mu_i = A_{ij}/A_{ji} \,,
\end{align}
the new system is reciprocal while the original one is NR, despite their identical dynamics.  Moreover, the new reciprocal system reaches thermal equilibrium if $\tilde{T}_i \equiv T$ with some $T > 0$, which, together with \eqref{condition2}, recovers~\eqref{eq:DB}. 

For a two-coordinate system, such a transformation exists if and only if $A_{12}A_{21}>0$ (take $\mu_1=|A_{12}|$, $\mu_2=|A_{21}|$).
In the general case, $A_{ij}A_{ji}>0$ is required for all $i,j$ (so antagonistic coupling is excluded), but this condition is not sufficient; for example there is no suitable transformation if $A$ has complex eigenvalues.
In such cases, NR interaction forces can be inferred by information-theoretic analysis of trajectories~\cite{hempel2024simple}.

There are also other cases where special forms of nonreciprocity retain or restore phenomenology usually associated with reciprocal or equilibrium dynamics. 
For example, ``perfectly NR couplings'' ($F_{ij}=F_{ji}$) can result in a Liouville-type theorem that enforces conservation of configuration-space volume and give rise to marginal orbits~\cite{hanai2024nonreciprocal}. Transverse NR forces can preserve the equilibrium Boltzmann distribution while sustaining circulating probability currents and finite entropy production~\cite{dopierala2026odd}. More generally, emergent symmetries can suppress manifestations of irreversibility: e.g., broken TRS  need not produce directed currents when an effective momentum-conservation law survives~\cite{metzger2026revisiting}. These examples emphasize that nonreciprocity does not uniquely determine the resulting nonequilibrium phenomenology: additional (effective) symmetries or conservation laws can strongly affect which signatures of irreversibility become observable.

\begin{figure*}
\centering
\includegraphics[width=15cm]{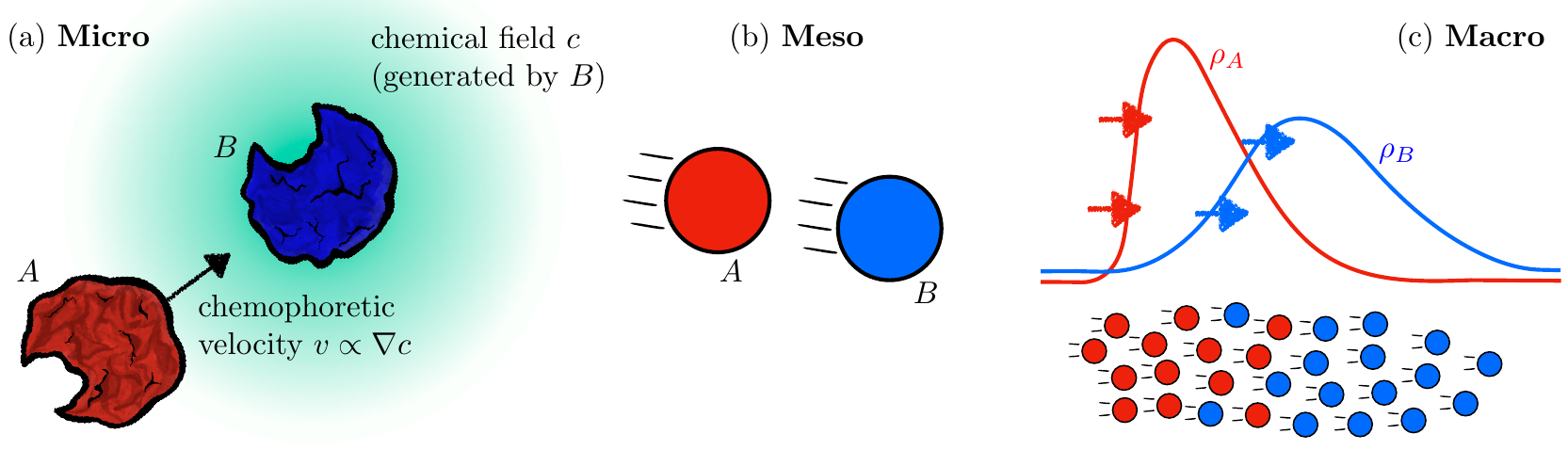}
\caption{\textbf{Nonreciprocity on different levels of description}.  (a)~Particle $A$ moves chemiphoretically towards particle $B$ (effective attraction) but the force on particle $B$ is only steric repulsion. (b)~If the phoretic effect is strong enough, integrating out microscopic details (e.g. the chemical field) yields a mesoscale model of particles with run-and-chase dynamics~\cite{jee2018catalytic,soto2014self,agudo2019active, dinelli2023non}. (c)~For many such particles, a macroscale description with continuous density fields supports travelling solutions, where species $A$ chases species $B$~\cite{saha2020scalar,you2020nonreciprocity}. The NR effective forces are indicated by arrows; they are accompanied by particle currents in the same direction~\cite{mason2025dynamical}.}  
\label{Fig2}
\end{figure*}

\section{Macroscale: Nonreciprocal Field-theories} 
We have argued that mesoscale nonreciprocity in physical systems can emerge from microscopically reciprocal dynamics through reduction. When large collections of such particles interact, a further coarse-graining step is required to describe their collective behavior in terms of continuum theories. Importantly, nonreciprocity need not survive this reduction. Conversely, macroscopic continuum models can themselves exhibit effective nonreciprocity, 
which however does not necessarily originate from underlying NR interactions: this parallels the emergence of NR mesoscale interactions from microscopic reciprocal dynamics.

To fix ideas, we focus on mixtures of interacting particles with overdamped dynamics.  On hydrodynamic length and time scales, this leads to continuity equations for the species' densities $\rho_a$:
\begin{align}
\dot\rho_a = -\nabla \cdot \Big( \mathbf{J}_a + \sum_b\sigma_{ab} \xi_b \Big)    \,,
\label{eq:rhoa-dot}
\end{align}
where $\mathbf{J}_a$ is the average current of species $a$ and the matrix $\sigma$ couples the density to independent unit spacetime white noises $\xi_b$.  This is a fluctuating hydrodynamic description that is obtained by coarse-graining on a scale much larger than any microscopic interaction range; the noise strength is small when the coarse-graining scale is large.

Constitutive relations give the dependence of $\bm{J}_a$ and $\sigma$ on the densities; these may be derived from underlying particle models (bottom-up approach) or proposed on general grounds that account for symmetries and conservation laws (top-down).
Bottom-up approaches often require approximations like a gradient expansion.  However macroscopic fluctuation theory~\cite{bertini2015} provides a route to derive such descriptions exactly, for idealised lattice models~\cite{mason2025dynamical}.

For mesoscale (particle) models, we emphasized the connection between nonreciprocity, TRS and momentum injection, and the concept of selfish energies. We will now discuss their analogies on the macroscale. To start, consider a generic fluctuating hydrodynamic theory in thermal equilibrium, whose currents can be factorised as
\beq
\textbf{J}_a = -\sum_b  M_{ab} \nabla \frac{\delta {\cal F}}{\delta \rho_b} \quad\text{with}\quad M_{ab}=\sum_k \sigma_{ak} \sigma_{bk}\,,
\label{eq:Jeq}
\eeq
where $\cal F$ is a free energy functional and $M$  is the mobility matrix, which is  positive-semidefinite and symmetric (Onsager reciprocity)~\cite{onsager1931}.
A convenient way to introduce NR interactions in this framework is to replace the free energy $\cal F$ by a species-dependent selfish free energy  ${\cal F}^{(a)}$:
\beq
\textbf{J}_a = -\sum_b M_{ab} \nabla \frac{\delta {\cal F}^{(a)}}{\delta \rho_b} \,.
\label{eq:Jeqself}
\eeq
Note that the mobility $M_{ab}$ is still symmetric: this setup includes NR interactions via selfish energies, but Onsager reciprocity is preserved for the mobility.

\newcommand{\nrs}{\nu}

\subsection{Example: Nonreciprocal Cahn-Hilliard Equation} 
A paradigmatic continuum description of conserved nonreciprocal systems is the NRCH~\cite{you2020nonreciprocity,saha2020scalar,brauns2024nonreciprocal,greve2025coexistence}. As an example, consider the density fluctuations $\rho_i=\bar\rho_i+\phi_i$ of two species $A,B$, driven by the currents
\begin{align}
\mathbf{J}_A & = (-\alpha - \phi_A^2 + \gamma \nabla^2)\nabla \phi_A - (\kappa-\nrs) \nabla \phi_B
\nonumber \\
\mathbf{J}_B&  = -\beta\nabla \phi_A - (\kappa+\nrs) \nabla \phi_A
\label{eq:nrch}
\end{align}
and conserved noise with $\sigma_{\alpha\beta}=\sqrt{2\epsilon}\delta_{\alpha\beta}$. 
While $\kappa$ controls reciprocal coupling, $\nrs$ introduces nonreciprocity. Antagonistic coupling appears if $|\nrs|>|\kappa|$.
The selfish free energies governing $A$ and $B$ in \eqref{eq:nrch} differ by $2\nrs\phi_A\phi_B/\epsilon$. 

The effect of nonreciprocity in \eqref{eq:nrch} appears clearly for $\alpha<0$ and $\beta>0$, where species $A$ would phase separate in isolation while species $B$ would remain homogeneous. For weak NR coupling, the system develops stationary structures (homogeneous or phase-separated). At the deterministic level, these states are indistinguishable from equilibrium phases, but their fluctuations break TRS for any $\nrs>0$.  Hence the entropy production rate is positive, which reveals the underlying nonequilibrium character~\cite{suchanek2023irreversible,suchanek2023entropy,alston2023irreversibility}.
Above a critical NR coupling, $\nrs^2>\beta^2+\kappa^2$, the system exhibits travelling-wave solutions in which one species effectively chases the other, providing the continuum analogue of run-and-chase dynamics (Fig.~\ref{Fig2}). The emergent travelling patterns clearly break TRS even in the deterministic limit~\cite{suchanek2023irreversible,suchanek2023entropy,alston2023irreversibility}.
The propagation direction of these waves is selected through spontaneous breaking of spatial inversion symmetry, resulting in finite integrated currents of both species~\cite{pisegna2024emergent}. This constitutes a key difference from mesoscale run-and-chase dynamics, where noise eventually reverses the propagation direction---reflecting the absence of true spontaneous symmetry breaking in finite particle systems.

\subsection{Momentum conservation}
For mesoscale systems, we have emphasized the connection between NR and momentum injection, providing another useful perspective on such emergent mass currents in overdamped systems. To discuss the macroscale analogy, we write a force-balance equation for species $a$. 
 This is a macroscopic analogue of the momentum balance equation~\eqref{eq:underdamped}, after taking the overdamped limit, spatial coarse-graining, and averaging over the noise. 
 Assuming dry friction and no external body forces (including no self-propulsion of the particles), one expects a general form~\cite{irving1950,wittkowski2017,zakine2020,metzger2026revisiting}
\beq
0 = \nabla\cdot \mathbf{\Pi}_a - \lambda_a \mathbf{J}_a +
\sum_{b(\neq a)} [ \mathbf{F}_{ab} - \Gamma_{ab} (\mathbf{v}_b-\mathbf{v}_a) ]
\label{eq:JF-bal}
\eeq
where $\mathbf{v}_a = \mathbf{J}_a/\rho_a$ is the drift velocity of species $a$,
 $\mathbf{\Pi}_a$ is the stress tensor for species $a$, $\lambda_a$ is its (dry) frictional coupling, $\Gamma$ is a matrix of (viscous) frictional couplings between the species, and $\mathbf{F}_{ab}$ is the remaining (non-frictional) force density on species $a$ from species $b$. 
 (Note that $\mathbf{\Pi}_a$, $\lambda_a$, $\mathbf{F}_{ab}$ and $\Gamma_{ab}$ generically depend on the local density and may also depend on gradients.)

Equation~\eqref{eq:JF-bal} is linear in the currents so it can be solved by matrix inversion to obtain $\mathbf{J}_a$.  For reciprocal systems one has $\mathbf{F}_{ab}=-\mathbf{F}_{ba}$ and $\Gamma_{ab}=\Gamma_{ba}$. 
Physically, this means that the dry friction still acts as a momentum sink and there is no possibility for injection of momentum, analogous to the situation on the particle level.  This implies $\sum_a \lambda_a \mathbf{J}_a = \nabla\cdot(\sum_a \mathbf{\Pi}_a)$ and integrating over a periodic domain yields zero on the RHS.  Noting that $\lambda_a>0$ for all species, this means that reciprocal systems cannot support run-and-chase situations in which all species have net mass currents that flow in the same direction.  
As noted above, travelling wave solutions in the NRCH have this property, which is a distinctive signature of nonreciprocity.


\textit{Linearised dynamics and bifurcations.} Another signature of NR interactions appears in the linearised dynamics about a homogeneous state.
Writing $\delta\rho_a$ for a small perturbation to $\rho_a$, the  behaviour on large length scales is given by an equation of the form
\beq
\partial_t (\delta \rho_a) = \nabla \cdot \Big[ \sum_b D_{ab} \nabla ( \delta \rho_b )  +  \sum_k \sigma_{ak} \xi_k \Big]\,.
\label{eq:lin-diff}
\eeq
For equilibrium dynamics as in Eq.~\eqref{eq:Jeq} one has $D_{ab}\propto\sum_c M_{ac} (\delta^2{\cal F}/\delta \rho_c \delta \rho_b)$
where $M$ is symmetric positive definite and the Hessian of $\mathcal F$ is symmetric; it follows that $D$ has real eigenvalues. 
However, the NR case of Eq.~\eqref{eq:Jeqself} allows $D$ to have complex eigenvalues. If the homogeneous state is stable then these manifest as oscillatory relaxation of density fluctuations, leading to the aforementioned breaking of TRS in the stationary state. If the nonreciprocity is strong enough, the homogeneous state can become unstable, via a Hopf-type bifurcation.
Such instabilities correspond to growing waves, which are also distinctive features of NR matter~\cite{you2020nonreciprocity,saha2020scalar}. They often result in travelling-wave states, although the eventual long-time dynamics depends on terms beyond the linear approximation of Eq.~\eqref{eq:lin-diff}.

We note in passing that a positive off-diagonal element $D_{ab}$ indicates that the current of species $a$ flows down gradients of $\rho_b$, which may suggest an effective repulsion between the species.  However, the sign of $D_{ab}$ is not sufficient to determine the sign of the interaction force in Eq.~\eqref{eq:Jeq}, which is $\nabla(\delta {\cal F}/\delta \rho_b)$: obtaining the current $\mathbf{J}$ and the diffusivity $D$ requires matrix multiplication by $M$.  In particular, reciprocal systems can have $D_{ab}$ and $D_{ba}$ with opposite signs if the matrix $M$ is not the identity (see e.g.~\cite{mason2025dynamical}).  On the other hand, NR couplings are necessary (but not sufficient) for $D$ to have complex eigenvalues.

Another hallmark of NR systems is the emergence of critical exceptional points~\cite{fruchart2021non,you2020nonreciprocity}, which can precede the emergence of spatiotemporal patterns such as traveling waves.  
{Unlike conventional critical points, they are governed not only by the spectrum but also by the geometry of the eigenvectors of the (non-Hermitian) stability operator, and are accompanied by strong fluctuations that markedly break TRS down to the deterministic limit~\cite{suchanek2023time}.}

\subsection{Nonreciprocity in active systems} 
Active matter provides an interesting context for disentangling TRS breaking and macroscale nonreciprocity~\cite{tevrugt2025exactly}. 
For a single active species without aligning interactions, a standard continuum description is Active Model B~\cite{wittkowski2014}. 
Taking $\rho=\bar\rho+\phi$, this is Eq.~\eqref{eq:rhoa-dot} with
\beq
\mathbf{J} = (-\alpha - \phi^2 + \gamma \nabla^2)\nabla \phi
- \lambda \nabla (|\nabla\phi|^2)
\label{eq:amb}
\eeq 
and fixed noise strength $\sigma$ (independent of $\phi$).
This is a nonequilibrium theory which breaks TRS and cannot be factorised as in Eq.~\eqref{eq:Jeq}. However, it is effectively reciprocal, which may be seen by the deterministic current remaining a pure gradient $\mathbf{J} = \nabla\cdot\mathbf{\Pi}$ with (total) isotropic stress $\mathbf{\Pi}$ proportional to the identity matrix. As above, integrating 
over a periodic domain shows that the total mass current vanishes.

Note that mesoscale (dry) active particle models typically include self-propulsion as an effective momentum source. Thus, the emergent reciprocity of Active Model B is a macroscale phenomenon which does not originate in a conservation law of the underlying particle description. (External potentials can generate persistent steady-state currents originating in the momentum injection via self-propulsion, which can be captured on the macroscopic scale by non-potential body forces~\cite{metzger2026revisiting}.)

In contrast, mixtures of active and passive species provide a striking example of effective NR interactions at the macroscale. Even when all particle-level interactions are reciprocal, 
coarse-graining yields NRCH-like equations with asymmetric couplings between the fields $\phi_A$ and $\phi_B$ representing the active and passive components, respectively~\cite{stenhammar2015activity,you2020nonreciprocity,mason2025dynamical}.
Unlike the single-species case, the momentum injected by the particles' self-propulsion generates macroscale nonreciprocity.  This in turn leads to
run-and-chase dynamics, Hopf bifurcations, travelling waves, and other pattern-forming solutions.
(The role of self-propulsion in generating macroscale nonreciprocity can be studied explicitly in
models where the active particle orientations are retained as slow variables and subsequently eliminated adiabatically~\cite{mason2025dynamical}.)
This is remarkable: adding a passive component to an active system is sufficient to break effective macroscale reciprocity---significantly broadening the 
scope of NR matter.

\section{Conclusions}
We have argued that nonreciprocity should be viewed as a physical property of some nonequilibrium systems, and that it is fundamentally scale-dependent: It may emerge or disappear through coarse-graining, and remain hidden or become apparent on different levels of description. Time-reversal symmetry provides a natural framework for tracking the manifestation of nonreciprocal behavior, linking it to observable nonequilibrium phenomena across scales.

This viewpoint raises several important questions for future work. Under which conditions does nonreciprocity survive coarse-graining and leave macroscopic traces? Which indicators of broken time-reversal symmetry remain robust across scales? 
Recently, progress on these questions was made in the vicinity of critical points, where long-range correlations dominate and such
questions can be addressed within the framework of the renormalization group~\cite{lorenzana2025nonreciprocity,akritidis2026fate,zelle2024universal,sezik2026critical,johnsrud2026scaling}.

A broader conceptual challenge is to delineate the scope of NR matter itself. Which emergent phenomena should be regarded as genuinely nonreciprocal, and which merely exhibit formal similarities? For example, flocking, Hall responses, or other non-Hermitian and nonequilibrium phenomena may share mathematical or physical signatures with NR systems, while originating from fundamentally different physical mechanisms. Establishing such distinctions will be essential for identifying universal principles of nonreciprocal matter and clarifying its place within the broader landscape of nonequilibrium statistical physics.

\acknowledgments
We thank 
Yael Avni,
Maria Bruna, 
Michael Cates,  
Erwin Frey, 
Michel Fruchart,
Ramin Golestanian,   
Ryo Hanai, 
Sabine Klapp,
Hartmut L\"owen,
Peter Sollich,
Anton Souslov,
Thomas Suchanek, 
Julien Tailleur, 
Uwe Thiele, and 
Vincenzo Vitelli 
for valuable discussions on nonreciprocity. We acknowledge support by grant NSF\,PHY-2309135 to the Kavli Institute for Theoretical Physics (KITP).

\bibliographystyle{iopart-num}
\bibliography{bib.bib} 

\end{document}